\documentclass[12pt]{article}
\usepackage{exscale,epsfig,a4}
\begin{document}
\title{{\bf Conditional quantum state engineering\\
       in repeated 2-photon down conversion}}
\author{J. Clausen$^1$,
        H. Hansen$^2$,
        L. Kn\"oll$^1$,
        J. Mlynek$^2$,
        D.--G. Welsch$^1$
\\[3ex]
$^1$ Friedrich-Schiller-Universit\"at Jena,\\
     Theoretisch-Physikalisches Institut,\\
     Max-Wien-Platz 1, D-07743 Jena, Germany,\\
     Fax: ++49 (0)3641 9 47102,\\
     Email: J.Clausen@tpi.uni-jena.de
\\[3ex]
$^2$ Universit\"at Konstanz,\\
     Fakult\"at f\"ur Physik,\\
     Universit\"atsstr. 10, D-78457 Konstanz, Germany,\\
     Fax: ++49 (0)7531 88-3072,\\
     Email: Hauke.Hansen@uni-konstanz.de
\\[1ex]
       } 
\date{}
\maketitle
\begin{abstract}
The U(1,1) and U(2) transformations realized by three-mode interaction in the 
respective parametric approximations are studied in conditional measurement, 
and the corresponding non-unitary transformation operators are derived.
As an application, the preparation of single-mode quantum states using an
optical feedback loop is discussed, with special emphasis of Fock state
preparation. For that example, the influence of non-perfect detection and
feedback is also considered.
\end{abstract}


\newpage
Conditional measurement offers a promising way to manipulate the state of a 
given quantum system. The basic idea is to entangle the state of the system 
under consideration with the state of an auxiliary system and to prepare the 
system in the desired state owing to the state reduction associated with an 
appropriate measurement on the auxiliary system. In what follows, we restrict 
our attention to travelling optical fields. The quantum states of two 
travelling modes can be entangled by mixing them at an appropriately chosen 
multiport. Possible basic transformations are the U(1,1) transformation as 
realized by a non-degenerate parametric amplifier and the U(2) transformation 
as realized by a frequency converter or a beam splitter.

The aim of this paper is to generalize and unify previous work on conditional 
measurement at U(2) and U(1,1) couplers. This includes the description of the 
quantum-state transformation in terms of a non-unitary operator \cite{Ban,pap2} 
as well as possible applications such as 
the generation of Fock states \cite{simultanFock}, 
Schr\"odinger cat-like states \cite{Anhut}, 
and photon-subtracted or photon-added Jacobi polynomial states \cite{Dakna}
or the measurement of specific overlaps \cite{phaseprosynth,pap4}.
By combining a theoretical concept for preparing single-mode quantum states 
by alternate coherent displacement and photon-adding \cite{pap3} with an 
experimental proposal to employ a 2-photon down converter inside a feedback 
loop \cite{Konstanz}, a way is offered to prepare quantum states without 
nonclassical input. In particular, emphasis is being placed on the generation 
of Fock states, for which possible imperfections of the optical components are 
analysed.

The paper is organized as follows.
In Sec.~\ref{sec1} appropriatly factorized representations of the unitary 
U(1,1) and U(2) transformation operators are introduced. They are used in 
Sec.~\ref{sec2} in order to derive the  nonunitary transformation operators 
realized by conditional measurement. In Sec.~\ref{sec3} the results are applied 
to the generation of specific quantum states, with special emphasis on Fock 
states and simple superpositions of Fock states. Finally, a summary and some 
concluding remarks are given in Sec.~\ref{sec4}. 
\section{U(1,1), U(2) and U($p$,$q$) from parametrically approximated 
         three-wave-mixing}
\label{sec1}
Let us consider the transformation
\begin{equation}
  \hat{\varrho}_{abc}^\prime
  =\hat{U}\hat{\varrho}_{abc}\hat{U}^\dagger
\label{1}
\end{equation}
of the quantum state $\hat{\varrho}_{abc}$ of three travelling optical mo\-des 
(denoted by $a$, $b$, and $c$), with 
$\hat{U}=\mathrm{e}^{-\frac{\mathrm{i}}{\hbar}\hat{H}t}$ being realized by a 
three-wave-mixer, 
\begin{equation}
  \hat{H}
  =\hbar\omega_a\hat{a}^\dagger\hat{a}
  +\hbar\omega_b\hat{b}^\dagger\hat{b}
  +\hbar\omega_c\hat{c}^\dagger\hat{c}
  +\hbar\chi^{(2)}
  (\hat{a}^\dagger\hat{b}^\dagger\hat{c}+\hat{c}^\dagger\hat{b}\hat{a}).
\label{H}
\end{equation}
Here, $t$ may be regarded as being the interaction time and $\chi^{(2)}$ 
corresponds to the second-order nonlinear susceptibility. If we assume that the 
mode $c$ is prepared in a coherent state, $\hat{\varrho}_{abc}$ $\!=$ 
$\!\hat{\varrho}_{ab}\otimes|\gamma\rangle\langle\gamma|$, then in the limit 
\mbox{$\chi^{(2)}$ $\!\to$ $\!0$} and $|\gamma|$ $\!\to$ $\!\infty$ with 
$\chi^{(2)}\gamma$ $\!=$ $\!$const. the reduced density operator 
$\hat{\varrho}_{ab}^\prime$ $\!=$
$\!\mathrm{Tr}_c\hat{\varrho}_{abc}^\prime$ reads 
\begin{equation}
  \hat{\varrho}_{ab}^\prime
  = \!\hat{U}_\mathrm{A}\hat{\varrho}_{ab}\hat{U}_\mathrm{A}^\dagger,
\label{3}
\end{equation}
where $\hat{U}_\mathrm{A}$ $\!=$
$\!\mathrm{e}^{-\frac{\mathrm{i}}{\hbar}\hat{H}_\mathrm{A}t}$
is a U(1,1) transformation realized by a non-degenerate parametric amplifier,
\begin{equation}
  \hat{H}_\mathrm{A}
  =\hbar\omega_a\hat{a}^\dagger\hat{a}
  +\hbar\omega_b\hat{b}^\dagger\hat{b}
  +\hbar\chi^{(2)}
  \big(\gamma\hat{a}^\dagger\hat{b}^\dagger+\gamma^*\hat{b}\hat{a}\big).
\label{4}
\end{equation}
Introducing the quantities
\begin{equation}
  \begin{array}{ll}
  \phi_0 = -(\omega_a-\omega_b)t,
  \quad&\quad
  \hat{K}_0 = \frac{1}{2}(\hat{a}\hat{a}^\dagger-
  \hat{b}^\dagger\hat{b}),\\
  \phi_1 = -(\gamma+\gamma^*)\chi^{(2)}t,
  \quad&\quad
  \hat{K}_1 = \frac{1}{2}(\hat{b}^\dagger\hat{a}^\dagger+
  \hat{a}\hat{b}),\\
  \phi_2 = -\mathrm{i}(\gamma-\gamma^*)\chi^{(2)}t,
  \quad&\quad
  \hat{K}_2 = \frac{1}{2\mathrm{i}}(\hat{b}^\dagger\hat{a}^\dagger-
  \hat{a}\hat{b}),\\
  \phi_3 = -(\omega_a+\omega_b)t,
  \quad&\quad
  \hat{K}_3 = \frac{1}{2}(\hat{a}\hat{a}^\dagger+
  \hat{b}^\dagger\hat{b}),
  \end{array}
\label{5}
\end{equation}
where the commutation relation
\begin{equation}
  [\hat{K}_j,\hat{K}_k]
  =\mathrm{i}\sum_{l,m=0}^3\varepsilon_{0jkl}g_{lm}\hat{K}_m
\label{6}
\end{equation}
is valid [$(g_{lm})$ $\!=$ $\!\mathrm{diag}(1,1,1,-1)$, $\varepsilon_{ijkl}$ 
is the four-di\-men\-sion\-al Levi-Civita symbol], and applying the respective 
disentanglement theorem \cite{factorization}, we may factorize 
$\hat{U}_\mathrm{A}$ as follows: 
\begin{eqnarray}
  \hat{U}_\mathrm{A}&=&
  \mathrm{e}^{-\frac{\mathrm{i}}{\hbar}\hat{H}_\mathrm{A}t}=
  \mathrm{e}^{\mathrm{i}\omega_at}\mathrm{e}^{\mathrm{i}(\phi_0\hat{K}_0
  +\phi_1\hat{K}_1+\phi_2\hat{K}_2+\phi_3\hat{K}_3)}\nonumber\\
  &=&
  \mathrm{e}^{-\mathrm{i}\frac{\phi_0+\phi_3}{2}}
  \mathrm{e}^{\mathrm{i}\phi_0\hat{K}_0}
  \mathrm{e}^{\mathrm{i}(\varphi_T + \varphi_R)\hat{K}_3}
  \mathrm{e}^{2\mathrm{i}\vartheta\hat{K}_2}
  \mathrm{e}^{\mathrm{i}(\varphi_T - \varphi_R)\hat{K}_3}\quad\nonumber\\
  &=& 
  \bar{T}^{*-1}{(PT)^*}^{-\hat{a}^\dagger\hat{a}}
  \mathrm{e}^{-P^*R\hat{b}^\dagger\hat{a}^\dagger}
  \mathrm{e}^{PR^*\hat{a}\hat{b}}
  {(PT^*)}^{-\hat{b}^\dagger\hat{b}},
\label{UA}
\end{eqnarray}
where
\begin{eqnarray}
  T&=&\cosh\vartheta \mathrm{e}^{\mathrm{i}\varphi_T}
  =\cosh\frac{\phi}{2}+\mathrm{i}\frac{\phi_3}{\phi}\sinh\frac{\phi}{2}\,,
  \label{TA}\\
  R&=&\sinh\vartheta\mathrm{e}^{\mathrm{i}\varphi_R}
  =\frac{\phi_2+\mathrm{i}\phi_1}{\phi}\sinh\frac{\phi}{2}\,,
  \label{RA}\\
  P&=&\mathrm{e}^{\mathrm{i}\frac{\phi_0}{2}},
\label{PA}
\end{eqnarray}
and $\bar{T}=T\mathrm{e}^{-\mathrm{i}\frac{\phi_3}{2}}$
($\phi$ $\!=$ $\!\sqrt{\phi_1^2+\phi_2^2-\phi_3^2}$).
For $\phi_0$ $\!=$ $\!0$ we have $P=1$ and $\hat{U}_\mathrm{A}$ in (\ref{UA}) 
reduces to a SU(1,1) transformation operator, compare \cite{interferometer}.

If, alternatively, we assume that the mode $b$ is prepared in a coherent state, 
$\hat{\varrho}_{abc}$ $\!=$
$\!\hat{\varrho}_{ac}\otimes|\beta\rangle\langle\beta|$, 
then in the limit $\chi^{(2)}$ $\!\to$ $\!0$ and
$|\beta|$ $\!\to$ $\!\infty$ with $\chi^{(2)}\beta$ $\!=$ $\!$const. the 
reduced density operator
$\hat{\varrho}_{ac}^\prime=\mathrm{Tr}_b\hat{\varrho}_{abc}^\prime$
reads
\begin{equation}
  \hat{\varrho}_{ac}^\prime
  =\hat{U}_\mathrm{C}\hat{\varrho}_{ac}\hat{U}_\mathrm{C}^\dagger,
\label{11}
\end{equation}
where $\hat{U}_\mathrm{C}$ $\!=$
$\!\mathrm{e}^{-\frac{\mathrm{i}}{\hbar}\hat{H}_\mathrm{C}t}$
is a U(2) transformation realized by a frequency converter,
\begin{equation}
  \hat{H}_\mathrm{C}
  =\hbar\omega_a\hat{a}^\dagger\hat{a}
  +\hbar\omega_c\hat{c}^\dagger\hat{c}
  +\hbar\chi^{(2)}
  \big(\beta\hat{c}^\dagger\hat{a}+\beta^*\hat{a}^\dagger\hat{c}\big).
\label{12}
\end{equation}
In order to factorize $\hat{U}_\mathrm{C}$, we introduce the quantities 
\begin{equation}
  \begin{array}{ll}
  \varphi_0 = -(\omega_a+\omega_c)t,
  \quad&\quad
  \hat{L}_0 = \frac{1}{2}(\hat{a}^\dagger\hat{a}+
  \hat{c}^\dagger\hat{c}),\\
  \varphi_1 = -(\beta^*+\beta)\chi^{(2)}t,
  \quad&\quad
  \hat{L}_1 = \frac{1}{2}(\hat{a}^\dagger\hat{c}+
  \hat{c}^\dagger\hat{a}),\\
  \varphi_2 = -\mathrm{i}(\beta^*-\beta)\chi^{(2)}t,
  \quad&\quad
  \hat{L}_2 = \frac{1}{2\mathrm{i}}(\hat{a}^\dagger\hat{c}-
  \hat{c}^\dagger\hat{a}),\\
  \varphi_3 = -(\omega_a-\omega_c)t,
  \quad&\quad
  \hat{L}_3 = \frac{1}{2}(\hat{a}^\dagger\hat{a}-
  \hat{c}^\dagger\hat{c}),
  \end{array}
\label{13}
\end{equation}
where now
\begin{equation}
  [\hat{L}_j,\hat{L}_k]
  =\mathrm{i}\sum_{l=0}^3\varepsilon_{0jkl}\hat{L}_l.
\label{14}
\end{equation}
We apply the respective
disentanglement theorem \cite{factorization} to obtain
\begin{eqnarray}
  \hat{U}_\mathrm{C}&=&
  \mathrm{e}^{-\frac{\mathrm{i}}{\hbar}\hat{H}_\mathrm{C}t}=
  \mathrm{e}^{\mathrm{i}(\varphi_0\hat{L}_0
  +\varphi_1\hat{L}_1+\varphi_2\hat{L}_2+\varphi_3\hat{L}_3)}\nonumber\\
  &=&
  \mathrm{e}^{\mathrm{i}\varphi_0\hat{L}_0}
  \mathrm{e}^{\mathrm{i}(\varphi_T + \varphi_R)\hat{L}_3}
  \mathrm{e}^{2\mathrm{i}\vartheta\hat{L}_2}
  \mathrm{e}^{\mathrm{i}(\varphi_T - \varphi_R)\hat{L}_3}\nonumber\\
  &=& 
  (\mathcal{PT})^{\hat{a}^\dagger\hat{a}}
  \mathrm{e}^{-\mathcal{PR}^*\hat{c}^\dagger\hat{a}}
  \mathrm{e}^{\mathcal{P^*R}\hat{a}^\dagger\hat{c}}
  (\mathcal{P^*T})^{-\hat{c}^\dagger\hat{c}},
\label{UC}
\end{eqnarray}
where  
\begin{eqnarray}
  \mathcal{T}&=&\cos\vartheta \mathrm{e}^{\mathrm{i}\varphi_T}
  =\cos\frac{\varphi}{2}+\mathrm{i}\frac{\varphi_3}{\varphi}
  \sin\frac{\varphi}{2}\,,
  \label{TC}\\
  \mathcal{R}&=&\sin\vartheta\mathrm{e}^{\mathrm{i}\varphi_R}
  =\frac{\varphi_2+\mathrm{i}\varphi_1}{\varphi}\sin\frac{\varphi}{2}\,,
  \label{RC}\\
  \mathcal{P}&=&\mathrm{e}^{\mathrm{i}\frac{\varphi_0}{2}}
\label{PC}
\end{eqnarray}
($\varphi$ $\!=$ $\!\sqrt{\varphi_1^2+\varphi_2^2+\varphi_3^2}$).
For $\varphi_0$ $\!=$ $\!0$ we have $\mathcal{P}=1$ and $\hat{U}_\mathrm{C}$ in 
(\ref{UC}) reduces to a SU(2) transformation operator, compare \cite{Campos}.

{F}rom (\ref{UA}) and (\ref{UC}) the transformation matrices 
for the respective mode operators are deduced to be
\begin{equation}
  \hat{U}_\mathrm{A}^\dagger{\hat{a} \choose \hat{b}^\dagger}
  \hat{U}_\mathrm{A}
  = P \left(\begin{array}{cc}
  T & -R \\
  -R^* & T^* 
  \end{array}\right)
  {\hat{a} \choose \hat{b}^\dagger}, 
  \label{MA}
\end{equation}
\begin{equation}
  \hat{U}_\mathrm{C}^\dagger{\hat{a} \choose \hat{c}}\hat{U}_\mathrm{C}
  = \mathcal{P} \left(\begin{array}{cc}
  \mathcal{T} & \mathcal{R} \\
  -\mathcal{R}^* & \mathcal{T}^* 
  \end{array}\right)
  {\hat{a} \choose \hat{c}}.
  \label{MC}
\end{equation}
In turn, (\ref{MA}) and (\ref{MC}) can themselves be used to define the 
four-parametric action of a parametric amplifier and a frequency converter.
In this case we consider $T$, $R$, $P$ and $\mathcal{T}$, $\mathcal{R}$, 
$\mathcal{P}$ as six complex numbers that satisfy the four conditions 
\begin{equation}
\label{21}
  |T|^2-|R|^2=|P|^2=|\mathcal{T}|^2+|\mathcal{R}|^2=|\mathcal{P}|^2=1
\end{equation}
and are otherwise arbitrary.
Using (\ref{MA}) and (\ref{MC}), it is not difficult to verify that
\begin{equation}
  [\hat{a}^\dagger\hat{a}-\hat{b}^\dagger\hat{b},\hat{U}_\mathrm{A}] =
  [\hat{a}^\dagger\hat{a}+\hat{c}^\dagger\hat{c},\hat{U}_\mathrm{C}] = 0.
\label{22}
\end{equation} 
The inverse transformations are obtained by replacing $\phi_j$ with $-\phi_j$ 
and $\varphi_j$ with $-\varphi_j$ in (\ref{TA}) -- (\ref{PA}) and
(\ref{TC}) -- (\ref{PC}), respectively, i.e.,
\begin{eqnarray}
  \hat{U}_\mathrm{A}^{-1}(T,R,P)&=&\hat{U}_\mathrm{A}(T^*,-R,P^*),
  \label{UAI}\\
  \hat{U}_\mathrm{C}^{-1}(\mathcal{T},\mathcal{R},\mathcal{P})
  &=&\hat{U}_\mathrm{C}(\mathcal{T}^*,-\mathcal{R},\mathcal{P}^*).
\label{UCI}
\end{eqnarray}
and interchanging signal and idler mode leads to
\begin{eqnarray}
  \hat{U}_\mathrm{A}(\hat{b},\hat{a};T,R,P)
  &=&\hat{U}_\mathrm{A}(\hat{a},\hat{b};T,R,P^*)
\label{25}\\
  \hat{U}_\mathrm{C}(\hat{c},\hat{a};\mathcal{T},\mathcal{R},\mathcal{P})
  &=&\hat{U}_\mathrm{C}(\hat{a},\hat{c};
  \mathcal{T}^*,-\mathcal{R}^*,\mathcal{P}).
\label{26}
\end{eqnarray}

If we assume that modes $b$ and $c$ are simultaneously 
prepared in coherent states, $\hat{\varrho}_{abc}$ $\!=$
$\!\hat{\varrho}_{a}\otimes|\beta\rangle\langle\beta|
\otimes|\gamma\rangle\langle\gamma|$, then in the limit 
$\chi^{(2)}$ $\!\to$ $\!0$, \mbox{$|\beta|$ $\!\to$ $\!\infty$},  
$|\gamma|$ $\!\to$ $\!\infty$ with $\chi^{(2)}\beta\gamma^*$ $\!=$ $\!$const. 
the reduced density operator $\hat{\varrho}_a^\prime$ $\!=$
$\!\mathrm{Tr}_{bc}\hat{\varrho}_{abc}^\prime$ reads 
$\hat{\varrho}_a^\prime$ $\!=$
$\!\hat{U}_\mathrm{D}\hat{\varrho}_a\hat{U}_\mathrm{D}^\dagger$, where
\begin{eqnarray}
  \hat{U}_\mathrm{D}&=&\mathrm{e}^{-\mathrm{i}[\omega_a\hat{a}^\dagger\hat{a}
  +\chi^{(2)}(\beta^*\gamma\hat{a}^\dagger+\beta\gamma^*\hat{a})]t}\nonumber\\
  &=&\mathrm{e}^{-\mathrm{i}\omega_at\hat{a}^\dagger\hat{a}}\hat{D}(\alpha)
\end{eqnarray}
can be written as the product of a U(1) transformation 
$\mathrm{e}^{-\mathrm{i}\omega_at\hat{a}^\dagger\hat{a}}$
and a coherent displacement $\hat{D}(\alpha$ $\!=$ $\!\mathrm{i}F^*)$
$\!=$ $\!\mathrm{e}^{\mathrm{i}\hat{F}}$, where
$\hat{F}$ $\!=$ $\!F\hat{a}$ $\!+$ $\!F^*\hat{a}^\dagger$ and 
$F$ $\!=$ $\!\mathrm{i}\chi^{(2)}\beta\gamma^*\omega_a^{-1}
(1$ $\!-$ $\!\mathrm{e}^{-\mathrm{i}\omega_at})$
(for details on the conditions under which parametric approximations hold, 
see \cite{D'Ariano}).

Linear coupling of more than two modes can be reduced to a successive 
application of two-mode couplers. For instance, as generalization of (\ref{MA}) 
and (\ref{MC}), let us consider a U($p$,$q$)-coupling of two sets of modes 
$a_1$, $\!\cdots$, $\!a_p$ and $a_{p+1}$, $\!\cdots$, $\!a_N$ 
\mbox{($N$ $\!=$ $\!p$ $\!+$ $\!q$)},
\begin{equation}
  \mathrm{e}^{-\mathrm{i}
  \hat{{\bf a}}^\dagger
  {\bf H}\hat{{\bf a}}}
  \hat{{\bf a}}
  \mathrm{e}^{\mathrm{i}
  \hat{{\bf a}}^\dagger
  {\bf H}\hat{{\bf a}}}
  =\mathrm{e}^{\mathrm{i}{\bf G}{\bf H}}\hat{{\bf a}},
\label{UNM}
\end{equation}
where $\hat{{\bf a}}$ is a column vector whose elements are
$\hat{a}_1$, $\!\cdots$, $\!\hat{a}_p$, $\!\hat{a}_{p+1}^\dagger$,
$\!\cdots$, $\!\hat{a}_{N}^\dagger$ and
$\hat{{\bf a}}^\dagger$ is a row vector with elements 
$\hat{a}_1^\dagger$, $\!\cdots$, $\!\hat{a}_p^\dagger$, $\!\hat{a}_{p+1}$,
$\!\cdots$, $\!\hat{a}_{N}$.
${\bf H}$ is a hermitian $N\times N$ matrix, and 
${\bf G}$ is a $N\times N$ diagonal matrix whose upper $p$ (lower $q$)
diagonal elements are equal to $1$ ($-1$) (see Appendix~\ref{app1}). 
Corresponding to a factorisation of 
$\mathrm{e}^{\mathrm{i}\hat{{\bf a}}^\dagger{\bf H}\hat{{\bf a}}}$, (\ref{UNM})
can be implemented by successive application of ${N \choose 2}$ two-mode 
couplers, each connecting two of the $N$ modes by either a U(2) transformation 
(if the two modes belong to the same set) or a U(1,1) transformation (if the 
two modes belong to different sets). The special case of implementing 
U($N$) $\!=$ $\!$U($N,0$) by means of beam splitters is discussed in 
\cite{unitary}. 
Note that $\mathrm{e}^{\mathrm{i}\hat{{\bf a}}^\dagger{\bf H}\hat{{\bf a}}}$
can alternatively be factorized into U(2) transformations and
single-mode squeezing operations \cite{Braunstein}.
\section{Conditional measurement at U(1,1) and U(2) couplers} 
\label{sec2}
\begin{figure}[htp]
\centerline{
\psfig{figure=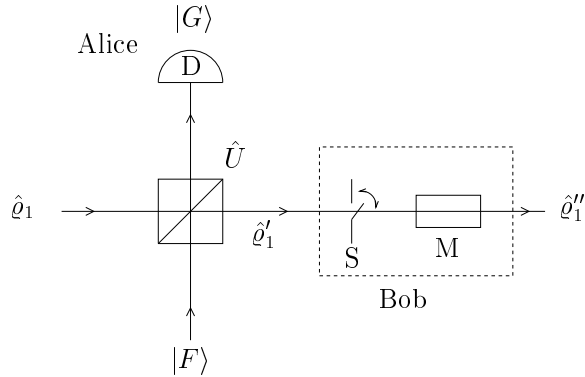,height=5cm}
}
\caption{
{\footnotesize
Schema of controlled quantum state engineering by conditional measurement at a 
U(1,1) or U(2) coupler realizing the transformation 
$\hat{U}$ $\!=$ $\!\hat{U}_\mathrm{A}$ or 
$\hat{U}$ $\!=$ $\!\hat{U}_\mathrm{C}$, respectively.
If Alice' measurement device  has detected a desired state $|G\rangle$, she 
informs Bob who opens the aperture S and 'stores' the pulse in the medium M 
until needed \protect\cite{Fleischhauer}.
}
\label{Fig1}
}
\end{figure}

Let us consider the scheme in Fig.~\ref{Fig1}. 
The signal mode $a$ (index $1$) prepared in a state $\hat{\varrho}_1$ and the 
idler mode ($b$ or $c$, index $2$) prepared in a state $|F\rangle$ are mixed at 
a U(1,1) parametric amplifier or a U(2) frequency converter, and a device D 
performs some measurement on the output idler mode. (The pump mode prepared in 
the strong coherent state is not shown in the figure.) Under the condition that 
D has detected a state $|G\rangle$, the reduced state of the output signal mode 
becomes
\begin{equation}
  \hat{\varrho}_1^\prime=\frac{1}{p}
  \mathrm{Tr}_2(\hat{U}\hat{\varrho}_1
  \otimes|F\rangle\langle F|\hat{U}^\dagger\hat{\Pi}),
\label{cond}
\end{equation}
where $\hat{\Pi}$ $\!=$ $\!|G\rangle\langle G|$. The norm \mbox{$p$ $\!=$
$\!\mathrm{Tr}_1\mathrm{Tr}_2$ $\!(\ldots)$} is the probability of measuring 
the state $|G\rangle$ and thus the probability of generating the state 
$\hat{\varrho}_1^\prime$. Introducing the non-unitary (conditional) operator
\begin{equation}
  \hat{Y}=\langle G|\hat{U}|F\rangle
\label{29}
\end{equation}
acting in the signal-mode Hilbert space, we can rewrite (\ref{cond}) as 
\begin{equation}
  \hat{\varrho}_1^\prime=\frac{1}{p}
  \hat{Y}\hat{\varrho}_1\hat{Y}^\dagger,
\label{conditionaltrafo}
\end{equation}
where the probability now reads $p$ $\!=$
$\!\mathrm{Tr}_1(\hat{Y}\hat{\varrho}_1\hat{Y}^\dagger)$.
In practice, synchronized sequences of light pulses could be fed into the input 
ports of the two-mode coupler, each pulse being prepared in the respective 
state, and the modes are thus non-monochromatic ones. In what follows we assume
that the U(1,1) and U(2) transformations do not vary with frequency within the 
spectral bandwidth of the pulses, so that the formulas given in Sec.~\ref{sec1} 
directly apply. 

In order to write $\hat{Y}$ in a closed form, we first represent the states 
$|F\rangle$ and $|G\rangle$ in the form of
\begin{eqnarray}
  |F\rangle&=&\hat{F}(\hat{b}^\dagger)|0\rangle
  =\sum_{m=0}^\infty F_m\hat{b}^{\dagger m}|0\rangle,
  \label{F}\\
  |G\rangle&=&\hat{G}(\hat{b}^\dagger)|0\rangle
  =\sum_{n=0}^\infty G_n\hat{b}^{\dagger n}|0\rangle
  \label{G}
\end{eqnarray}
(or with $\hat{b}^\dagger$ being replaced by $\hat{c}^\dagger$) and substitute 
these expressions into (\ref{29}). Using (\ref{UA}) and (\ref{UC}), applying 
the $s$-ordering rule \cite{Cahill}
\begin{eqnarray}
  \{\hat{a}^{\dagger m}\hat{a}^n\}_s
  &=&\sum_{k=0}^{\min[m,n]}
  k!{m \choose k}{n \choose k}\left(\frac{t-s}{2}\right)^k\nonumber\\
  &&\times\,\{\hat{a}^{\dagger m-k}\hat{a}^{n-k}\}_t,
\label{ordering}
\end{eqnarray}
and introducing $\hat{n}$ $\!=$ $\!\hat{a}^\dagger\hat{a}$,
we obtain (see Appendix~\ref{app2}) 
\begin{eqnarray}
  \hat{Y}_\mathrm{A} 
  &\hspace{-.2ex}=&\hspace{-.2ex}
  \langle0|\hat{G}^\dagger(\hat{b}^\dagger)\hat{U}_\mathrm{A}
  \hat{F}(\hat{b}^\dagger)|0\rangle\nonumber\\
  &\hspace{-.2ex}=&\hspace{-.2ex}
  \left\{\hat{G}^\dagger\left(-R^*T^{-1}\hat{a}\right)
  \hat{F}(P^*R^*\hat{a})
  \right\}_{s_\mathrm{A}}\!\!\!\bar{T}^{*-1}{(PT)^*}^{-\hat{n}},\qquad
  \label{YA}\\
  \hat{Y}_\mathrm{C} 
  &\hspace{-.2ex}=&\hspace{-.2ex}
  \langle0|\hat{G}^\dagger(\hat{c}^\dagger)\hat{U}_\mathrm{C}
  \hat{F}(\hat{c}^\dagger)|0\rangle\nonumber\\
  &\hspace{-.2ex}=&\hspace{-.2ex}\left\{
  \hat{G}^\dagger\left(-\mathcal{R}\mathcal{T}^{*-1}\hat{a}^\dagger\right)
   \hat{F}(\mathcal{P}\mathcal{R}\hat{a}^\dagger)
  \right\}_{s_\mathrm{C}}\!\!\!(\mathcal{P}\mathcal{T})^{\hat{n}},
  \label{YC}
\end{eqnarray}
where the respective ordering parameters are given by
\begin{equation}
  s_\mathrm{A}=\left|\frac{|T|^2+1}{|T|^2-1}\right|,\quad
  s_\mathrm{C}=\left|\frac{|\mathcal{T}|^2+1}{|\mathcal{T}|^2-1}\right|.
\label{36}
\end{equation}
Since $|\mathcal{T}|^2$ $\!\le$ $\!1$ $\!\le$ $\!|T|^2$, we see that 
$s_\mathrm{A},s_\mathrm{C}$ $\!\ge$ $\!1$. 
Note that the arguments of $\hat{F}$ and $\hat{G}$ in (\ref{YC}) are the 
adjoints of the corresponding arguments in (\ref{YA}).

Coherent displacements can be separated from the ordering procedure. To see
this, we derive from (\ref{MA}) and (\ref{MC}) together with (\ref{UAI}) and
(\ref{UCI}) the transformation formulas for the displacement operators of the 
signal and idler mode,
\begin{eqnarray}
  \lefteqn{
  \hat{U}_\mathrm{A}\;\hat{D}_1(\alpha)\hat{D}_2(\beta)\;
  \hat{U}_\mathrm{A}^\dagger}\nonumber\\
  &&\hspace{1ex}
  =\hat{D}_1[P(T\alpha-R\beta^*)]
  \hat{D}_2[P^*(-R\alpha^*+T\beta)],
\label{DA}
\end{eqnarray}
\begin{eqnarray}
  \lefteqn{  
  \hat{U}_\mathrm{C}\;\hat{D}_1(\alpha)
  \hat{D}_2(\beta)\;
  \hat{U}_\mathrm{C}^\dagger}\nonumber\\
  &&\hspace{1ex}
  =\hat{D}_1[\mathcal{P}(\mathcal{T}\alpha+\mathcal{R}\beta\;)]
  \hat{D}_2[\mathcal{P}(-\mathcal{R}^*\alpha+\mathcal{T}^*\beta)].
\label{DC}
\end{eqnarray}
Combining (\ref{DA}) and (\ref{DC}) with (\ref{YA}) and (\ref{YC}), 
respectively, yields
\begin{eqnarray} 
  \hat{Y}_\mathrm{A}^\prime 
  &=&\langle0|\hat{G}^\dagger(\hat{b}^\dagger)
  \hat{D}_2^\dagger(\beta)\hat{U}_\mathrm{A}\hat{D}_2(\alpha)
  \hat{F}(\hat{b}^\dagger)|0\rangle\nonumber\\
  &=& \hat{D}_1\!\!
  \left(\frac{P\alpha^*-T\beta^*}{R^*}\right)
  \hat{Y}_\mathrm{A}
  \hat{D}_1\!\!\left(\frac{P^*\beta^*-T^*\alpha^*}{R^*}\right),\quad
  \label{DYA} \\
  \hat{Y}_\mathrm{C}^\prime 
  &=&\langle0|\hat{G}^\dagger(\hat{c}^\dagger)
  \hat{D}_2^\dagger(\beta)\hat{U}_\mathrm{C}\hat{D}_2(\alpha)
  \hat{F}(\hat{c}^\dagger)|0\rangle\nonumber\\
  &=& \hat{D}_1\!\!
  \left(\frac{\mathcal{P}\alpha-\mathcal{T}\beta}{\mathcal{R}^*}\right)
  \hat{Y}_\mathrm{C}
  \hat{D}_1\!\!\left(
  \frac{\mathcal{P}^*\beta-\mathcal{T}^*\alpha}{\mathcal{R}^*}\right),
  \label{DYC}
\end{eqnarray} 
i.e., a coherent displacement of the idler mode is equivalent to a 
corresponding coherent displacement of the signal mode. 

Since each trial in Fig.~\ref{Fig1} yields a desired measurement outcome only 
with some probabilility, a lockable aperture S is needed in order to extract 
the properly transformed outgoing signal modes and dump the others. The desired 
output states are thus available at random times. It may however be demanded to 
provide them at certain times. In this case, a quantum state memory M has to be 
used into which the pulses can be fed and released when desired. One
possibility to realize M is offered by electromagnetically induced transparency 
\cite{Fleischhauer}.
If, in particular, a pulse train of a certain repetition frequency is required,
one may apply, e.g., an array of delays with variable optical path lengths 
which step by step adjust the waiting periods between the pulses to each other.
Besides electromagnetically induced transparency, the cross-Kerr effect should 
also offer a way to realize a variable optical delay. 
In the latter case the outgoing signal mode in the state 
$\hat{\varrho}_1^\prime$ is mixed with a reference mode (index 3) 
prepared in a coherent state $|\alpha\rangle$ at a cross-Kerr medium 
$\hat{U}$ $\!=$ $\!\mathrm{e}^{\mathrm{i}\kappa\hat{n}_1\hat{n}_3}$. 
For $\kappa$ $\!\to$ $\!0$ and $|\alpha|$ $\!\to$ $\!\infty$ with
$\kappa|\alpha|^2$ $\!=$ $\!$const., the reduced signal state becomes 
$\hat{\varrho}_1^{\prime\prime}$ $\!=$ $\!\mathrm{Tr}_3(
\hat{U}\hat{\varrho}_1^\prime|\alpha\rangle\langle\alpha|\hat{U}^\dagger
)$ $\!=$ $\!\mathrm{e}^{\mathrm{i}\kappa|\alpha|^2\hat{n}_1}
\hat{\varrho}_1^\prime\mathrm{e}^{-\mathrm{i}\kappa|\alpha|^2\hat{n}_1}$.
Via choosing $|\alpha|^2$ we can therefore control the refractive index and
with it the optical length and time delay caused by the cross-Kerr medium.
\section{Preparation of single-mode quantum states}
\label{sec3}
\subsection{Displaced photon adding for generating qubits}
\label{sec3.1}
\begin{figure}[htp]
\centerline{
\psfig{figure=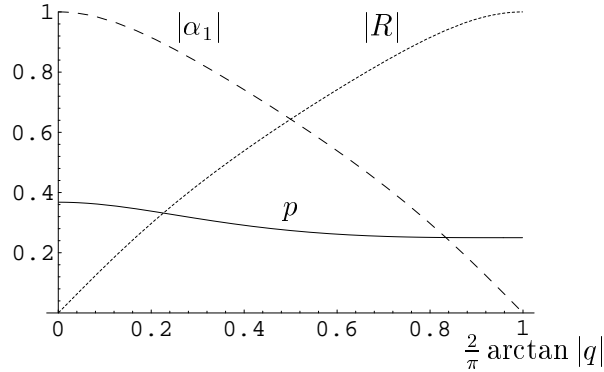,height=5cm}
}
\caption{
{\footnotesize
The maximized probability (\protect\ref{p}) of generating a qubit 
(\protect\ref{qubit}) and the corresponding values of the parameters 
$|R|$ and $|\alpha_1|$ as functions of $|q|$.
}
\label{Fig2}
}
\end{figure}

In order to illustrate the general results derived in Sec.~\ref{sec2}, let us 
first study the generation of a single qubit. 
When a parametric amplifier is fed with an idler pulse prepared in a 
single-mode coherent state $|F\rangle$ $\!=$ $\!|\alpha_1\rangle$ and a single 
photon is detected, $|G\rangle$ $\!=$ $\!|1\rangle$, then (\ref{YA}) and 
(\ref{DYA}) yield the non-unitary (conditional) operator
\begin{eqnarray}
  \lefteqn{
  \hat{Y}^{(k)}_\mathrm{A} = -RP^*\bar{T}^{*-1}
  \hat{D}\!\left(PR^{*-1}\alpha^*_k\right)
  }\nonumber\\
  &&\hspace{6ex}\times\,
  (PT)^{*-\hat{n}}\hat{a}^\dagger
  \hat{D}\!\left(-T^*R^{*-1}\alpha^*_k\right)
\label{Yk}
\end{eqnarray}
(the index $k$ $\!=$ $\!1$ is introduced for later purposes).
Let us further assume that the signal input channel is unused,
$\hat{\varrho}$ $\!=$ $\!|0\rangle\langle0|$
(for notational convenience we omit the mode index).  
The outgoing signal pulse is then prepared in a state
\begin{equation}
  \hat{\varrho}^\prime=\frac{1}{p}
  \hat{Y}^{(1)}_\mathrm{A}|0\rangle
  \langle0|\hat{Y}^{(1)}_\mathrm{A}{^\dagger}
  =|\Psi\rangle\langle\Psi|,
\label{42}
\end{equation}
where
\begin{equation}
  |\Psi\rangle=\frac{|0\rangle+q|1\rangle}{\sqrt{1+|q|^2}}
\label{qubit}
\end{equation}
is a superposition of the vacuum and a single-photon Fock state. 
The parameter $q$ $\!=$ $\!-PR/\alpha_1$ can be controlled
by varying $R$ or $\alpha_1$. 
It is however convenient to choose $|R|$ and $|\alpha_1|$ such that for a 
desired $q$ the probability
\begin{equation}
  p=\|\hat{Y}^{(1)}_\mathrm{A}|0\rangle\|^2
  =(|R|^2+|\alpha_1|^2)|T|^{-4}\mathrm{e}^{-|\alpha_1|^2}
\label{p}
\end{equation}
$\big(\||\Phi\rangle\|$ $\!=$ $\!\sqrt{\langle\Phi|\Phi\rangle}\big)$
of generating the qubit (\ref{qubit}) attains a maximum. This is the case for
\begin{equation}
  |R|^2=\frac{\sqrt{(|q|^{-2}+1)^2+4|q|^{-2}}-(|q|^{-2}+1)}{2|q|^{-2}}.
\label{popt}
\end{equation}
The maximum values of $p$ together with the corresponding values
of $|R|$ and $|\alpha_1|$ are shown in Fig.~\ref{Fig2}.
\subsection{Repeated displaced photon adding for generating
            arbitrary superpositions of Fock states}
\label{sec3.2}
\begin{figure}[htp]
\centerline{
\psfig{figure=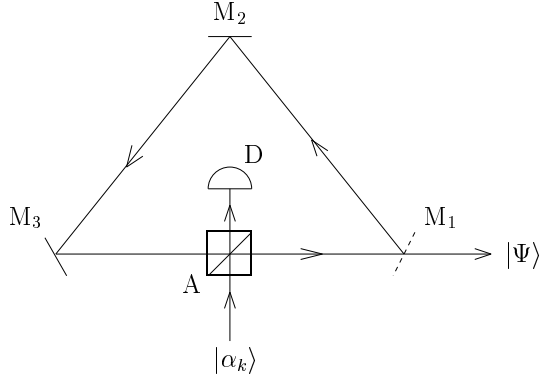,height=5cm}
}
\caption{
{\footnotesize
Scheme for preparing a travelling optical field in a quantum state (\ref{fs}). 
A parametric amplifier A is fed with a sequence of idler pulses prepared in 
appropriately chosen coherent states $|\alpha_k\rangle$. 
They (and the pump pulses, not shown) arrive at A simultaneously with the 
produced signal pulse circulating in the ring resonator consisting of mirrors 
$\mathrm{M}_2$, $\mathrm{M}_3$ and a (removable) mirror $\mathrm{M}_1$. The 
desired quantum state is generated if in each round trip of the signal field 
the detector D registers a single photon.
}
\label{Fig3}
}
\end{figure}

The scheme can be extended to the generation of 
an arbitrary superposition of $N$ Fock states
\begin{equation}
  |\Psi\rangle=\sum_{n=0}^N|n\rangle\langle n|\Psi\rangle.
\label{fs}
\end{equation}
Since the states of that type are completely determined by the $N$ zeros
of the $Q$-function, i.e., the $N$ solutions $\beta_1,\cdots,\beta_N$
of the equation $\langle\Psi|\beta\rangle$ $\!=$ $\!0$, they can be
generated from the vacuum by alternate application of the coherent
displacement operator and the creation operator,
\begin{eqnarray}
  |\Psi\rangle&=&\frac{\langle N|\Psi\rangle}{\sqrt{N!}}\prod_{k=1}^N
  (\hat{a}^\dagger-\beta_k^*)|0\rangle\nonumber\\
  &=&\frac{\langle N|\Psi\rangle}{\sqrt{N!}}\prod_{k=1}^N\left[
  \hat{D}(\beta_k)\hat{a}^\dagger\hat{D}^\dagger(\beta_k)\right]|0\rangle,
\label{sollstate}
\end{eqnarray}
which may be realized repeating the procedure in Sec.~\ref{sec3.1} 
according to Fig.~\ref{Fig3}.
The pulse prepared in the state (\ref{qubit}) is sent back through a ring 
resonator to the amplifier and used as signal input. Simultaneously,
an idler pulse prepared in a coherent state $|\alpha_2\rangle$ is fed into the 
second input port of the amplifier. If the detector registers again a single 
photon, then the outgoing signal pulse is prepared in a state 
$\sim\hat{Y}^{(2)}_\mathrm{A}\hat{Y}^{(1)}_\mathrm{A}|0\rangle$. 
Under the condition that in each round trip a single photon is detected, after 
$N$ round trips the signal pulse is prepared in a state 
\begin{equation}
  |\Psi\rangle \sim \hat{Y}_\mathrm{A}|0\rangle
  =\hat{Y}^{(N)}_\mathrm{A}\cdots
  \hat{Y}^{(2)}_\mathrm{A}\hat{Y}^{(1)}_\mathrm{A}|0\rangle.
\label{Yoverall}
\end{equation}
If the preparation of the desired state has been successful, the mirror 
$\mathrm{M}_1$ in Fig.~\ref{Fig3} can be removed in order to open the cavity 
and release the pulse. Otherwise the pulse is dumped in order to start the next 
trial from the very beginning, with the signal input port of the amplifier 
being unused.

Inserting (\ref{Yk}) and rearranging the operator order such that the photon 
creation operators are on the left of the exponential operators, we derive
\begin{eqnarray}
\lefteqn{
  \hat{Y}_\mathrm{A}|0\rangle =
  \mathrm{e}^{\mathrm{i}\xi}|R|^N|T|^{-\frac{N}{2}(N+3)}
  \,\mathrm{exp}\!\!\left(-\frac{1}{2}\sum_{k=1}^N|\alpha_k|^2\right)
  }
  \nonumber\\&&\hspace{2ex}\times
  \prod_{k=1}^N\left[\hat{a}^\dagger-\frac{(PT)^{*N}}{R}\sum_{l=k}^N
  \frac{P^*\alpha_l-T\alpha_{l+1}}{(PT)^{*l}}\right]|0\rangle
\label{iststate}
\end{eqnarray}
($\alpha_{N+1}$ $\!=$ $\!0$; $\mathrm{e}^{\mathrm{i}\xi}$
irrelevant phase factor). In the derivation of (\ref{iststate}) we have used 
the relations
$\hat{D}(\alpha)f(\hat{a},\hat{a}^\dagger)\hat{D}^\dagger(\alpha)$
$\!=$ $\!f(\hat{a}-\alpha,\hat{a}^\dagger-\alpha^*)$ and
$\alpha^{\hat{n}}f(\hat{a},\hat{a}^\dagger)\alpha^{-\hat{n}}$
$\!=$ $\!f(\alpha^{-1}\hat{a},\alpha\hat{a}^\dagger)$.
Comparing (\ref{iststate}) with (\ref{sollstate}), we see that for 
\begin{equation}
  \beta_k=\frac{(PT)^N}{R^*}
  \sum_{l=k}^N\frac{P\alpha_l^*-T^*\alpha_{l+1}^*}{(PT)^l}
\label{50}
\end{equation}
or equivalently
\begin{equation}
  \alpha_k=\frac{PR}{(PT)^k(PT)^{*N}}
  \sum_{l=k}^N|T|^{2l}(\beta_l^*-\beta_{l+1}^*)
\label{51}
\end{equation}
($\beta_{N+1}$ $\!=$ $\!0$) the desired state $|\Psi\rangle$ is just realized. 
The probability of generating the state is given by
\begin{eqnarray}
  p&=&\|\hat{Y}_\mathrm{A}|0\rangle\|^2\nonumber\\
  &=&\frac{N!}{|\langle N|\Psi\rangle|^2}
  \frac{|R|^{2N}}{|T|^{N(N+3)}}\,
  \mathrm{exp}\!\!\left(-\sum_{k=1}^N|\alpha_k|^2\right),
\label{probability}
\end{eqnarray}
as is seen by comparing the norms of both sides of (\ref{iststate}) and 
inserting (\ref{sollstate}). 
It decreases rapidly with increasing $N$ in general, so that the applicability 
of this method is effectively restricted to low numbers of round trips.
\subsection{Fock state generation}
\label{sec3.3}
In order to generate a Fock state $|n\rangle$, no feeding with idler modes is 
necessary, since all the $\beta_k$ in (\ref{sollstate}) are zero. According to 
(\ref{probability}), the probability of detecting one photon at each of the $n$ 
round trips is
\begin{equation}
  p = n!|R|^{2n}|T|^{-n(n+3)}
\label{probabilityfock}
\end{equation}
and becomes maximal for fixed $n$ if $|R|^2$ $\!=$ $\!2/(n$ $\!+$ $\!1)$. 
Substituting this expression into (\ref{probabilityfock}) and applying 
Stirling's formula yields the asymptotic behaviour
$p_\mathrm{max}$ $\!\approx$ $\!a\mathrm{e}^{-bn}$, where
$a$ $\!=$ $\!\sqrt{2\pi}/\mathrm{e}$ and
$b$ $\!=$ $\!2-\ln2$, i.e., the probability of preparing an $n$-photon Fock 
state  decreases exponentially with increasing $n$. Note that exactly the same 
asymptotic behaviour is observed when the state is generated by conditional 
measurement at a beam splitter array \cite{pap3}. 

However, the scheme considered here offers the specific feature of
circumventing the problem of low preparation probability. Since the idler field 
remains in the vacuum state, $|\alpha_k\rangle$ $\!=$ $|0\rangle$, the
circulating signal pulse is always in a Fock state whose number is simply the 
sum of detected idler photons. This means that every trial sooner or later 
results in the desired state $|n\rangle$, provided that the respective $n$ is 
not skipped, and thus there is no need to wait for a sequence of $n$ successive
single-photon clicks. The idler detector is simply used for photon book-keeping 
and the cavity is opened in that moment when the sum of all detected idler 
photons has reached the desired value $n$.
\subsection{Influence of non-perfect photodetection and non-per\-fect 
            cavity feedback on Fock state generation}
\label{sec3.4}
If the idler detector and the feedback mirrors are not perfect, the situation 
becomes more complicated. Let us assume that the state of the signal pulse 
after the $j$th round trip is a statistical mixture of Fock states described by 
a density operator $\hat{\varrho}^{(j)}$. This pulse now enters the signal 
input port of the parametric amplifier whose idler input port is unused, 
$|F\rangle$ $\!=$ $\!|0\rangle$. When $k$ outgoing idler photons are detected 
with efficiency $\eta_\mathrm{D}$, then the state $\hat{\varrho}^{(j)\prime}$ 
of the outgoing signal field is given by (\ref{cond}), where now
\begin{equation}
  \hat{\Pi}(k) = \;:
  \frac{\left(\eta_\mathrm{D}\hat{n}\right)^k}{k!}
  \,\mathrm{e}^{-\eta_\mathrm{D}\hat{n}}:\;
  =\hat{b}_{k\hat{n}}(\eta_\mathrm{D}).
\label{POM}
\end{equation}
Here, the symbol $:\;:$ introduces normal ordering, and 
\begin{equation}
  {b}_{kn}(z)={n \choose k}z^k(1-z)^{n-k}.
\label{55}
\end{equation}
Inserting (\ref{POM}) into (\ref{cond}), we find that the prepared state is a 
mixture of Fock states with
\begin{equation}
  \varrho^{(j)\prime}_{mm}=\frac{1}{|T|^2 p(k)}\sum_l
  b_{k,m-l}(\eta_\mathrm{D})b_{lm}(|T|^{-2})
  \varrho^{(j)}_{ll},
\label{pmprime}
\end{equation}
where
\begin{equation}
  p(k)=|T|^{-2}\sum_{l,m}b_{k,m-l}(\eta_\mathrm{D})
  b_{lm}(|T|^{-2})\varrho^{(j)}_{ll}
\label{mixeddistribution}
\end{equation}
is the probability of detecting $k$ photons. Next, the pulse prepared in the 
state $\hat{\varrho}^{(j)\prime}$ is fed back by the mirrors 
$\mathrm{M}_1,\cdots,\mathrm{M}_3$ into the signal input port of the amplifier. 
If the reflectances of the mirrors $\mathrm{M}_1,\mathrm{M}_2,\mathrm{M}_3$ are 
given by $\mathcal{R}_1,\mathcal{R}_2,\mathcal{R}_3$, the state of the pulse 
after the $(j$ $\!+$ $\!1)$th round trip is still a mixture of Fock states with
\begin{equation}
  \varrho^{(j+1)}_{mm}
  =\sum_l b_{ml}(\eta_\mathrm{F})\varrho^{(j)\prime}_{ll},
\label{passingLC}
\end{equation}
where $\eta_\mathrm{F}=|\mathcal{R}_1\mathcal{R}_2\mathcal{R}_3|^2$ determines 
the feedback efficiency. Combining (\ref{passingLC}) and (\ref{pmprime}), we 
obtain the recursion relation 
\begin{eqnarray}
  \lefteqn{
  \varrho^{(j+1)}_{mm}
  =\frac{1}{|T|^2 p(k_{j+1})}
  \sum_{l,n}b_{ml}(\eta_\mathrm{F})
  }\nonumber\\
  &&\hspace{8ex}\times\,
  b_{k_{j+1},l-n}(\eta_\mathrm{D})
  b_{nl}(|T|^{-2})
  \varrho^{(j)}_{nn}.
\label{staterecursion}
\end{eqnarray}
If we start from the vacuum state, $\varrho^{(0)}_{nn}$ $\!=$ $\!\delta_{n0}$, 
and measure the numbers $k_{j}$ of (outgoing) idler photons detected at the 
$j$th round trip, the circulating pulse therefore evolves into a mixture of 
Fock states, which depends according to (\ref{staterecursion}) on the 
respective sequence $\{k_1,k_2,\cdots\}$. For this reason, some arbitrariness 
has to be introduced if the state evolution wants to be simulated.

Let us therefore first consider the evolution of the {\it mean} photon number 
of the circulating pulse which can be obtained by considering the case when no 
measurement is performed in the idler output channel. 
(\ref{staterecursion}) together with $\eta_\mathrm{D}$ $\!=$ $\!0$ and 
\mbox{$k_{j+1}$ $\!=$ $\!0$} 
(and the initial condition $\varrho^{(0)}_{nn}$ $\!=$ $\!\delta_{n0}$) 
then yields a thermal state
\begin{equation}
  \hat{\varrho}^{(N)}
  =\frac{1}{\langle\hat{n}\rangle^{(N)}+1}\,
  \hat{b}_{0\hat{n}}\!\!
  \left(\frac{1}{\langle\hat{n}\rangle^{(N)}+1}\right),
\label{thermal}
\end{equation}
where the mean photon number
\begin{equation}
  \langle\hat{n}\rangle^{(N)}=\eta_\mathrm{F}|R|^2
  \frac{\left(\eta_\mathrm{F}|T|^2\right){^{\!N}}-1}{\eta_\mathrm{F}|T|^2-1}
\label{thermalmean}
\end{equation}
can be deduced from the recursion relation
\begin{equation}
  \langle\hat{n}\rangle^{(j+1)}
  =\eta_\mathrm{F}\left[|T|^2\langle\hat{n}\rangle^{(j)}+|R|^2\right].
\label{62}
\end{equation}
If $\eta_\mathrm{F}|T|^2$ $\!>$ $\!1$, then $\langle\hat{n}\rangle^{(N)}$
increases exponentially with $N$, while for $\eta_\mathrm{F}|T|^2$ $\!<$ $\!1$ 
a stationary value is observed, $\lim_{N\to\infty}\langle\hat{n}\rangle$ 
$\!=$ $\!\eta_\mathrm{F}|R|^2/(1-\eta_\mathrm{F}|T|^2)$.
The critical value $\eta_\mathrm{F}|T|^2$ $\!=$ $\!1$ leads to 
a linear increase, $\langle\hat{n}\rangle^{(N)}$ $\!=$ $\!|R/T|^2N$.

\begin{figure}[htp]
\centerline{
\psfig{figure=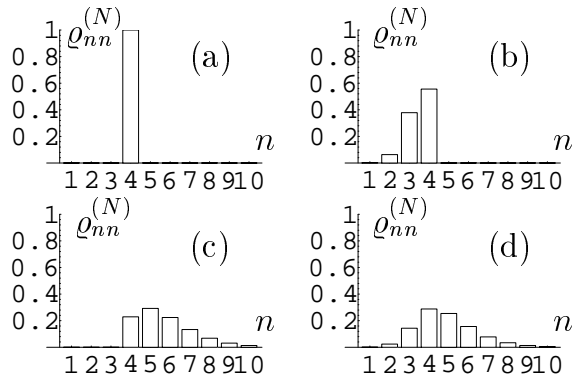,height=5cm}
}
\caption{
{\footnotesize
Computer simulation of the preparation of a desired Fock state 
$|n$ $\!=$ $\!4\rangle$. The density matrix elements $\varrho^{(N)}_{nn}$ 
obtained according to (\ref{staterecursion}) together with 
$\varrho^{(0)}_{nn}$ $\!=$ $\!\delta_{n0}$ are shown for different 
feedback efficiencies $\eta_\mathrm{F}$ and detection efficiencies 
$\eta_\mathrm{D}$, and $|R|^2$ $\!=$ $\!3\cdot 10^{-3}$.
The cavity is assumed to be opened after detecting the 4th idler photon, which 
determines the number of round trips $N$.
\protect\\
\protect\begin{tabular}{ll}
(a) $\eta_\mathrm{F}=1$ and $\eta_\mathrm{D}=1$ ($N=538$). &
(b) $\eta_\mathrm{F}=0.999$ and $\eta_\mathrm{D}=1$ ($N=652$).
\protect\\
(c) $\eta_\mathrm{F}=1$ and $\eta_\mathrm{D}=0.7$ ($N=636$). &
(d) $\eta_\mathrm{F}=0.999$ and $\eta_\mathrm{D}=0.7$ ($N=788$).
\protect\end{tabular}
}
\label{Fig4}
}
\end{figure}

To give an example, let us now consider the generation of the state 
$|n$ $\!=$ $\!4\rangle$. Since the amplification is typically weak 
($|R|^2\ll1$), a large number of round trips is likely and a high-quality 
cavity is demanded. Note that $|R|^2$ $\!=$ 
$\!_s\langle\mathrm{vac}|\hat{a}^\dagger\hat{a}|\mathrm{vac}\rangle_s$
$\!=$ $\!_s\langle\mathrm{vac}|\hat{b}^\dagger\hat{b}|\mathrm{vac}\rangle_s$
characterizes the mean photon numbers of a two-mode squeezed vacuum
$|\mathrm{vac}\rangle_s$ $\!=$ $\!\hat{U}_\mathrm{A}|0,0\rangle$
that is generated by means of a parametric amplifier.
In order to provide a rough estimate of the required efficiency 
$\eta_\mathrm{F}$, we estimate the number $N$ of cycles needed on average by 
identifying $\langle\hat{n}\rangle{^{(N)}}$ in (\ref{thermalmean}) with $n$ and 
assume a linear increase $\langle\hat{n}\rangle{^{(N)}}$ $\!=$ $\!|R/T|^2N$
(because of the smallness of $|R|^2$), so that $N$ $\!\approx$ $\!|R|^{-2}n$. 
We now estimate $\eta_\mathrm{F}$ from the requirement that after that number 
of round trips in an empty resonator (i.e., without amplifier) an initially 
present photon can still be found with probability $1/2$, thus 
$\eta_\mathrm{F}$ $\!=$ $\!2^{-|R|^2/n}$.
For $n$ $\!=$ $\!4$ and $|R|^2$ $\!=$ $\!3\cdot 10^{-3}$ this yields 
$\eta_\mathrm{F}\approx0.999$.
We insert these quantities into (\ref{thermalmean}) and (arbitrarily)
consider the case when the first idler photon is detected if 
$\eta_\mathrm{F}\langle\hat{n}\rangle$ exceeds 1, the second if 
$\eta_\mathrm{F}\langle\hat{n}\rangle$ exceeds 2 and so on,
until eventually the cavity is opened after detecting the 4th idler photon. 
The resulting density matrix is then obtained from (\ref{staterecursion}). 
Examples are shown in Fig.~\ref{Fig4}. The plots confirm the sensitivity to 
cavity loss for small $|R|^2$. Note that the Mandel-parameter
$Q$ $\!=$ $\!\langle(\Delta\hat{n})^2\rangle/\langle\hat{n}\rangle-1$
of the mixture plotted in (d) is $Q$ $\!=$ $\!-0.527$.
\section{Conclusion}
\label{sec4}
In this paper we have studied conditional quantum-state engineering at 
parametric amplifiers and frequency converters, regarding each apparatus
as being effectively a two-port device, whose action in conditional measurement 
can be described by a non-unitary operator $\hat{Y}$ defined in the Hilbert 
space of the signal mode. We have presented $\hat{Y}$ for arbitrary quantum 
states of the incoming idler mode and arbitrary detected quantum states of the 
outgoing idler mode as $s$-ordered products of the operators that generate the 
quantum states from the vacuum, $s$ being entirely determined by the devise 
parameters. 

To illustrate the results, we have proposed a scheme allowing the generation of 
arbitrary finite single-mode quantum states of travelling waves by a parametric 
amplifier equipped with a ring resonator as an optical feedback loop. We have 
applied the method to the problem of Fock state preparation, for which we have 
also addressed the influence of non-perfect photodetection and non-perfect 
cavity feedback.
\section*{Acknowledgements}
This work was supported by the Deutsche Forschungsgemeinschaft.
\begin{appendix}
\section{Derivation of Equation (\protect\ref{UNM})}
\label{app1}
The commutation relation between $\hat{{\bf a}}$ and $\hat{{\bf a}}^\dagger$
is invariant under a U($p$,$q$) transformation, i.e., for
${\bf U}^{-1}={\bf G}^\dagger{\bf U}^\dagger{\bf G}$ we have
\begin{equation}
  [({\bf U}\hat{{\bf a}})_\lambda,({\bf U}\hat{{\bf a}})_\mu^\dagger]
  =[\hat{{\bf a}}_\lambda,\hat{{\bf a}}_\mu^\dagger]={\bf G}_{\lambda\mu}.
\label{commu}
\end{equation}
After writing
\begin{eqnarray}
  \hat{{\bf a}}\mathrm{e}^{\mathrm{i}\hat{{\bf a}}^\dagger{\bf H}\hat{{\bf a}}}
  &=&{\bf U}^{-1}{\bf U}\hat{{\bf a}}
  \mathrm{e}^{\mathrm{i}\hat{{\bf a}}^\dagger
  {\bf U}^\dagger{\bf U}^{\dagger-1}{\bf H}{\bf U}^{-1}{\bf U}\hat{{\bf a}}}
  \nonumber\\
  &=&{\bf U}^{-1}({\bf U}\hat{{\bf a}})
  \mathrm{e}^{\mathrm{i}({\bf U}\hat{{\bf a}})^\dagger
  ({\bf U}^{-1\dagger}{\bf H}{\bf U}^{-1})({\bf U}\hat{{\bf a}})},
\label{eq64}
\end{eqnarray}
we choose ${\bf U}$ such that 
${\bf U}^{-1\dagger}{\bf H}{\bf U}^{-1}$ becomes diagonal, i.e.,
$({\bf U}^{-1\dagger}{\bf H}{\bf U}^{-1})_{\lambda\mu}$
$\!=$ $\!({\bf U}^{-1\dagger}{\bf H}{\bf U}^{-1})_{\lambda\lambda}
\delta_{\lambda\mu}$.
Applying the relation $\alpha^{\hat{n}}f(\hat{a},
\hat{a}^\dagger)\alpha^{-\hat{n}}$ 
$\!=$ $\!f(\alpha^{-1}\hat{a},\alpha\hat{a}^\dagger)$,
we see that the relation
$\hat{{\bf a}}_\mu\mathrm{e}^{\mathrm{i}\alpha
\hat{{\bf a}}_\lambda^\dagger\hat{{\bf a}}_\lambda}$
$\!=$ $\!\mathrm{e}^{\mathrm{i}\alpha
\hat{{\bf a}}_\lambda^\dagger\hat{{\bf a}}_\lambda}
\mathrm{e}^{\mathrm{i}{\bf G}_{\lambda\mu}\alpha}\hat{{\bf a}}_\mu$ holds.
Inserting it into (\ref{eq64}) with 
$\alpha$ $\!=$ $\!({\bf U}^{-1\dagger}{\bf H}{\bf U}^{-1})_{\lambda\lambda}$, 
we obtain together with (\ref{commu})
\begin{eqnarray}
  \hat{{\bf a}}\mathrm{e}^{\mathrm{i}\hat{{\bf a}}^\dagger{\bf H}\hat{{\bf a}}}
  &=&{\bf U}^{-1}({\bf U}\hat{{\bf a}})
  \prod_\lambda\mathrm{e}^{\mathrm{i}
  ({\bf U}^{-1\dagger}{\bf H}{\bf U}^{-1})_{\lambda\lambda}
  ({\bf U}\hat{{\bf a}})_\lambda^\dagger
  ({\bf U}\hat{{\bf a}})_\lambda}\nonumber\\
  &=&{\bf U}^{-1}
  \mathrm{e}^{\mathrm{i}\hat{{\bf a}}^\dagger{\bf H}\hat{{\bf a}}}
  \mathrm{e}^{\mathrm{i}{\bf G}({\bf U}^{-1\dagger}{\bf H}{\bf U}^{-1})}
  ({\bf U}\hat{{\bf a}})\nonumber\\
  &=&\mathrm{e}^{\mathrm{i}\hat{{\bf a}}^\dagger{\bf H}\hat{{\bf a}}}
  {\bf U}^{-1}
  \mathrm{e}^{\mathrm{i}{\bf G}{\bf U}^{-1\dagger}{\bf H}{\bf U}^{-1}}
  {\bf U}\hat{{\bf a}}\nonumber\\
  &=&\mathrm{e}^{\mathrm{i}\hat{{\bf a}}^\dagger{\bf H}\hat{{\bf a}}}
  \mathrm{e}^{\mathrm{i}{\bf G}{\bf H}}\hat{{\bf a}}.
\end{eqnarray}
Note that $\mathrm{e}^{\mathrm{i}{\bf G}{\bf H}}$ is just a U($p$,$q$) 
matrix, because $\mathrm{e}^{-\mathrm{i}{\bf G}{\bf H}}
={\bf G}^\dagger\mathrm{e}^{-\mathrm{i}{\bf H}^\dagger{\bf G}^\dagger}{\bf G}$.
\section{Derivation of Equation (\protect\ref{YA})}
\label{app2}
Applying (\ref{MA}), we first write
\begin{equation}
  \hat{b}^n\hat{U}_\mathrm{A}=\hat{U}_\mathrm{A}
  (\hat{U}_\mathrm{A}^\dagger\hat{b}^\dagger\hat{U}_\mathrm{A})^{\dagger n}
  =\hat{U}_\mathrm{A}(-P^*R\hat{a}^\dagger+P^*T\hat{b})^n.
\label{lemma1}
\end{equation}
Using
$\alpha^{\hat{n}}f(\hat{a},\hat{a}^\dagger)\alpha^{-\hat{n}}$
$\!=$ $\!f(\alpha^{-1}\hat{a},\alpha\hat{a}^\dagger)$,
we see from (\ref{UA}) that for $|0\rangle\equiv|0\rangle_2$
\begin{equation}
  \langle0|\hat{U}_\mathrm{A}
  =\bar{T}^{*-1}(PT)^{*-\hat{a}^\dagger\hat{a}}
  \langle0|\mathrm{e}^{\frac{R^*}{T^*}\hat{a}\hat{b}}.
\label{lemma2}
\end{equation}
With the help of (\ref{lemma1}) and (\ref{lemma2}) we now get
\begin{eqnarray}
  \hat{Y}_\mathrm{A}
  &=&\sum_{m,n} F_m G_n^*
  \langle0|\hat{b}^n\hat{U}_\mathrm{A}\hat{b}^{\dagger m}|0\rangle
 \nonumber\\
  &=&\sum_{m,n,k} F_m G_n^*
  \bar{T}^{*-1}(PT)^{*-\hat{a}^\dagger\hat{a}}
  \langle0|\mathrm{e}^{\frac{R^*}{T^*}\hat{a}\hat{b}}\nonumber\\
  &&\times{n \choose k}(P^*T\hat{b})^k(-P^*R\hat{a}^\dagger)^{n-k}
  \hat{b}^{\dagger m}|0\rangle
 \nonumber\\
  &=&\sum_{m,n,k} F_m G_n^*
  \bar{T}^{*-1}(PT)^{*-\hat{a}^\dagger\hat{a}}
  k!{m \choose k}{n \choose k}(P^*T)^k\nonumber\\
  &&\times(R^*T^{*-1}\hat{a})^{m-k}(-P^*R\hat{a}^\dagger)^{n-k}
 \nonumber\\
  &=&\sum_{m,n,k} F_m G_n^*
  \bar{T}^{*-1}
  k!{m \choose k}{n \choose k}(P^*T)^k\nonumber\\
  &&\times(P^*R^*\hat{a})^{m-k}(-RT^{*-1}\hat{a}^\dagger)^{n-k}
  (PT)^{*-\hat{n}}
 \nonumber\\
  &=&\sum_{m,n} F_m G_n^* (P^*R^*)^m (-RT^{*-1})^n
  \nonumber\\
  &&\times\sum_k
  k!{m \choose k}{n \choose k}\left(-\left|\frac{T}{R}\right|^2\right)^k
  \hat{a}^{m-k}\hat{a}^{\dagger n-k}
  \nonumber\\
  &&\times\bar{T}^{*-1}(PT)^{*-\hat{n}}.
\end{eqnarray}
Applying (\ref{ordering}) with $t=-1$ (denoting antinormal order) and
$s=s_\mathrm{A}$ gives 
\begin{eqnarray}
  \hat{Y}_\mathrm{A}
  &=&\sum_{m,n} F_m G_n^* (P^*R^*)^m (-RT^{*-1})^n
  \nonumber\\
  &&\times\{\hat{a}^{\dagger n}\hat{a}^m\}_{s_\mathrm{A}}
  \bar{T}^{*-1}(PT)^{*-\hat{n}}
 \nonumber\\
  &=&\left\{\sum_{m,n} G_n^* (-R^*T^{-1}\hat{a})^{\dagger n} 
  F_m (P^*R^*\hat{a})^m\right\}_{s_\mathrm{A}}
  \nonumber\\
  &&\times\bar{T}^{*-1}(PT)^{*-\hat{n}},
\end{eqnarray}
from which (\ref{YA}) directly follows. (\ref{YC}) can be derived analogously.
\end{appendix}

\end{document}